# Ultra-Efficient Kidney Stone Fragment Removal via Spinner-Induced Synergistic Circulation and Spiral Flow

Yilong Chang[1], Jasmine Vallejo[1], Yangqing Sun[1], Ruike Renee Zhao[1] *

[1]Department of Mechanical Engineering, Stanford University; Stanford, CA 94305, USA.

*Corresponding author. Email: rrzhao@stanford.edu

**Abstract:**

Kidney stones can cause severe pain and complications such as chronic kidney disease or kidney failure. Retrograde intrarenal surgery (RIRS), which uses laser lithotripsy to fragment stones for removal via a ureteroscope, is widely adopted due to its safety and effectiveness. However, conventional fragment removal methods using basketing and vacuum-assisted aspiration are inefficient, as they can capture only 1–3 fragments (1–3 mm in size) per pass, often requiring dozens to hundreds of ureteroscope passes during a single procedure to completely remove the fragments. These limitations lead to prolonged procedures and residual fragments that contribute to high recurrence rates. To address these limitations, we present a novel spinner device that enables ultra-efficient fragment removal through spinning-induced localized suction. The spinner generates a three-dimensional spiral and circulating flow field that dislodges and draws fragments into its cavity even from distances up to 20 mm, eliminating the need to chase fragments. It can capture ~60 fragments (0.5–2 mm) or ~15 larger fragments (2–3 mm) in a single pass, significantly improving removal efficiency. In this work, the spinner design is optimized via computational fluid dynamics to maximize suction performance. In-vitro testing demonstrates near 100% capture rates for up to 60 fragments in a single operation and superior large distance capture efficacy compared to vacuum-assisted methods. Ex-vivo testing of the integrated spinner–ureteroscope system in a porcine kidney confirmed its high performance by capturing 45 fragments in just 4 seconds during a single pass and achieving complete fragment clearance within a few passes. The spinner demonstrates its high potential to dramatically improve procedural efficiency by reducing operative time, minimizing ureteroscope passes, and enhancing stone-free rates.

**Introduction**

Kidney stones are mineral deposits in the renal system that cause severe pain, bleeding, and complications like chronic kidney disease or life-threatening kidney failure(*1-3*). Among the available treatment strategies, retrograde intrarenal surgery (RIRS), a minimal invasive procedure that uses a laser to fragment kidney stones for removal, is most commonly recommended due to its high effectiveness and safety(*4, 5*). During RIRS, a ureteroscope is inserted through the urinary tract to access the kidney, allowing direct visualization of the kidney stone (**Fig. 1A**) (*6*). Laser lithotripsy is then performed, breaking large stones into small fragments (2-4 mm (*7*)) or submillimeter dust (<1 mm (*8, 9*)). To enhance stone clearance, RIRS often incorporates basketing, a technique in which a nitinol basket retrieves the stone fragments one by one (**Fig. 1B**) (*10*). Although basketing achieves a higher reported stone free rate (SFR) (defined as number of patient who achieve stone free out of total patient population, and stone free is defined as no residual stone fragments) of approximately 78% (*11, 12*), it requires excessive ureteroscope passes, from dozens to hundreds during a single procedure, making it extremely time-consuming and inefficient.

More recently, a newer technique using vacuum-assisted dedusting lithotripsy demonstrates enhanced stone removal efficiency and a higher SFR (*13-16*) by aspirating stone dust (<1 mm) during or after laser lithotripsy (*17, 18*)(**Fig. 1B**). However, larger fragments (1~3 mm), which constitute the majority of the fragmented stone, still require manual retrieval, leading to repeated ureteroscope insertions that significantly slow the procedure and limit the efficiency of the technique (*19*). Additionally, effective suction relies heavily on precise fragment/dust targeting and close proximity, which complicates the procedure and limits the operation efficiency. Despite these recent advancements, up to 50% of patients experience kidney stone recurrence in part due to residual fragments (*20, 21*), leading to costly retreatments and increased burdens on patients, physicians, and healthcare systems (*22*). Thus, the development of new stone removal mechanisms and technologies that can enhance stone removal efficiency, reduce operation time, and achieve a high SFR remains an urgent need.

To address these challenges, in this work, we develop a novel kidney stone collection technology that utilizes spinning-induced localized suction to enhance the stone removal efficiency. This innovative approach enables the rapid capture and collection of large number of stone fragments in a single ureteroscope pass, significantly improving stone clearance speed. For 0.5~2 mm fragments, the spinner can collect ~60 fragments or ~15 fragments (2~3 mm) per pass, compared to basketing and vacuum-assisted methods, which usually remove one to three fragments at a time. This dramatic increase in capture capacity can substantially reduce procedure time. As shown in **Fig. 1C**, the spinner mechanism utilizes a rotating component that generates a three-dimensional rotating and circulating flow to effectively dislodge and capture stone fragments. This unique flow dynamics creates localized suction within the spinner cavity, enabling the rapid attraction and capture of stone fragments for efficient and effective stone removal. Beyond its large capture capacity, the spinner can effectively draw in stone fragments from distances up to 20 mm, eliminating the need to chase fragments, which is a limitation in basketing and vacuum-assisted techniques. This further accelerates the stone removal process, enhancing overall efficiency of the procedure.

In the following study, we first quantitively investigate the structural design of the spinner to optimize its suction performance for enhanced kidney stone fragments capturing through computational (CFD) simulation. With the optimized design, the spinner's fragments capture ability is evaluated against different quantities of gravel fragments, demonstrating near 100% capture rate for 40, 50, and 60 fragments in a single operation. Additionally, the far-distance capture capability of the spinner is characterized, showing a greater than 90% capture rate for gravel fragments positioned between 2 to 15 mm away. Comparison test on fragments capture rate is conducted against vacuum-assisted technique with the fragments placed at various distances to demonstrate spinner's superior fragments capture capability. Then, *in-vitro* fragment removal is performed in a kidney flow model demonstrating efficient removal of 40 gravel fragments in a single operation, achieving a complete clearance in 40 seconds. Finally, the spinner is integrated with the existing ureteroscope to perform *ex-vivo* fragments removal in porcine kidney. The spinner-ureteroscope system demonstrates high efficiency, capturing 45 fragments within 4 seconds of spinning and removing them in a single ureteroscope pass. Ultimately, the spinner successfully clears fragments from both the renal pelvis and the upper pole.

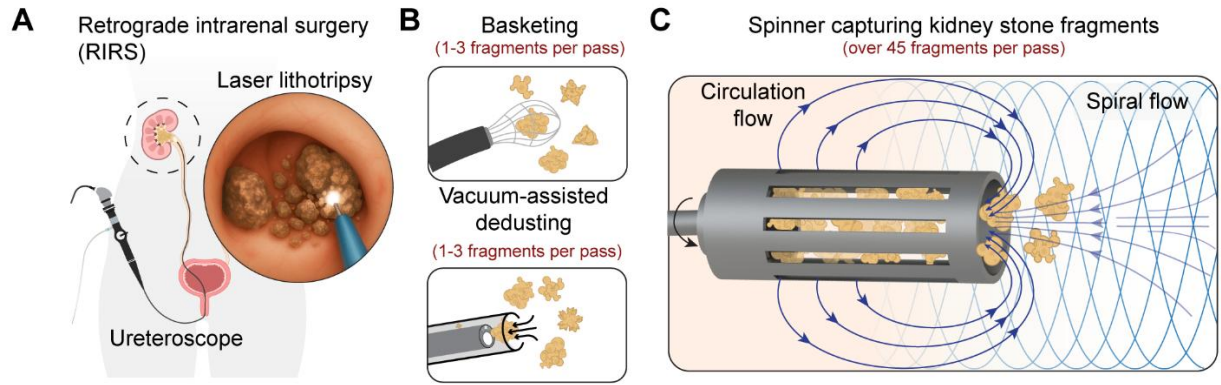

**Fig. 1. Spinner mechanism for ultra-efficient kidney stone fragment capture and removal.** (**A**) Schematic of retrograde intrarenal surgery (RIRS) procedure illustrating fragmentation of a large kidney stone. (**B**) Schematic of conventional fragment removal techniques during RIRS: basketing and vacuum-assisted dedusting. (**C**) Schematic of the spinner mechanism generating 3D circulation and spiral flow to dislodge, draw in, and capture stone fragments.

Results

**Spinner structural optimization for enhanced fragment capturing through localized suction**

The spinner's effectiveness in capturing kidney stone fragments from a distance is driven by a strong localized suction force, which is induced from a dual-rotation flow mechanism that dislodge and draw the fragments into the spinner. This flow field arises from the spinner's unique geometry, a hollow cylinder (sealed at one end) with slits along its wall. The key dimensions of the spinner with $ID$ =3 mm are indicated in **Fig. 2A**. The working principle of spinner is illustrated by CFD simulations (See Materials and Methods for simulation details). The 3D streamline and velocity field of the spinner rotating at 10k rpm in a 15mm-diamater tube is illustrated in **Fig. 2B.** The spinner's rotation creates a 3D flow field, where circulation flow and spiral flow synergistically enhance stone fragment capture. The slits on the spinner wall facilitate a circulation flow, producing strong localized suction that is evident in the high flow velocity inside the spinner cavity (**Fig. 2C**). A zoomed-in view at the bottom of **Fig. 2C** clearly depicts the flow entering through the spinner opening and exiting through the slit, effectively drawing and collecting stone fragments into the cavity. In parallel, the spinner's rotation induces a spiral flow field, as visualized in the velocity contour with streamlines for selected tubular cross-sections along the spinner axis (**Fig. 2D**). The rotational flow patterns observed at various planes ($z_1$-$z_3$) extending 15 mm from the spinner demonstrate the ability to shear and dislodge stone fragments from a distance. This transition from static to dynamic motion ensures that the fragments follow the generated flow field, guiding them efficiently into the spinner cavity for effective capture.

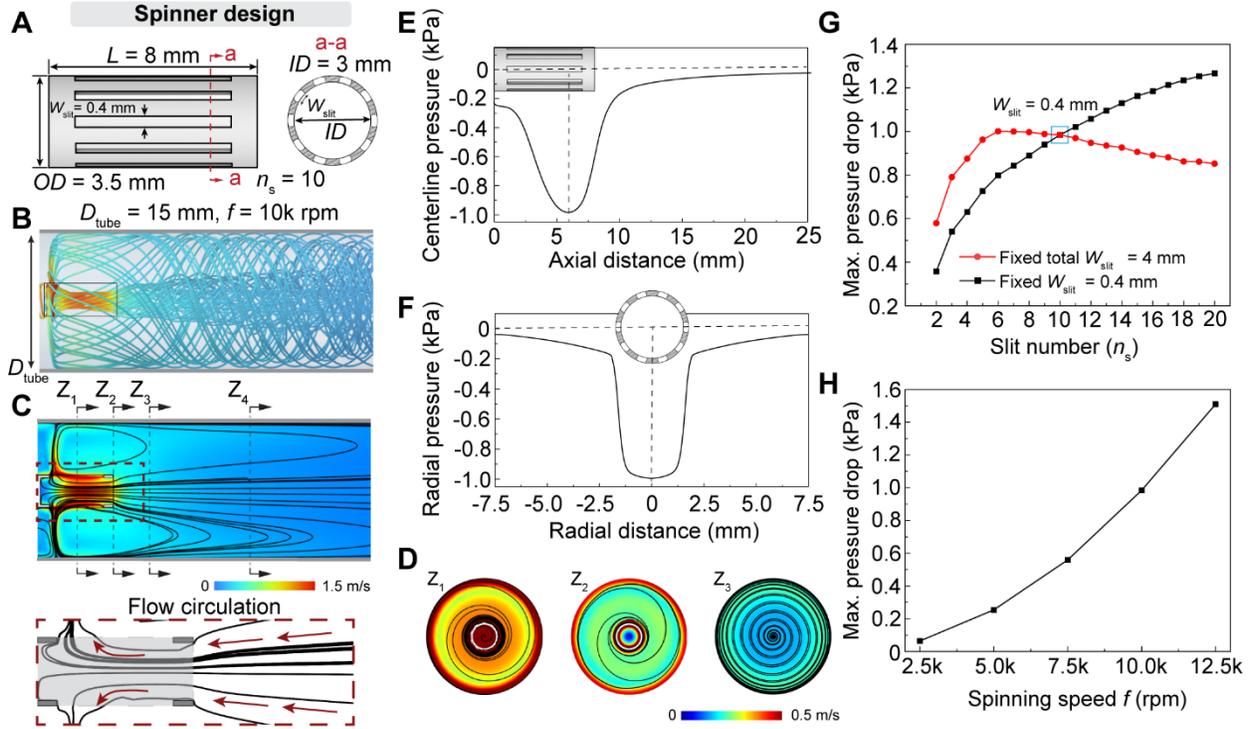

**Fig. 2. Stone capturing mechanism and structural optimization of the spinner for enhanced suction.** (**A**) Schematic of spinner design, with length $L = 8$ mm, outer diameter $OD = 3.5$ mm, inner diameter $ID = 3$ mm and 10 evenly spaced slits around the cylinder. Each slit has a width of $W_{slit} = 0.4$ mm. (**B**) 3D streamlines generated by the spinner at 10k rpm inside a 15 mm diameter tube (approximating renal pelvis size) showing rotational flow patterns. (**C**) Velocity contour and streamlines along the spinner's longitudinal section, illustrating circulation flow. The zoomed-in view depicts fluid entering the spinner front and exiting through the slits. (**D**) Velocity contours and streamlines at three planes ($z_1 = -4$ mm, $z_2 = 0$ mm, $z_3 = 15$ mm) relative to the spinner's front surface. (**E**) Centerline pressure distribution along spinner axis showing a sharp pressure drop inside spinner cavity for suction. (**F**) Radial pressure distribution across the spinner cross-section at the location of maximum axial pressure drop. (**G**) Maximum centerline pressure drop with respect to slit number $n_s$ at 10k rpm (spinner $ID = 3$ mm, $W_{slit} = 0.4$ mm and total $W_{slit} = 4$ mm). (**H**) Maximum centerline pressure drop at varying spinning frequency (2.5k, 5.0k, 7.5k, 10k, and 12.5k rpm).

Beyond facilitating circulation flow for collecting stone fragments, the slits on the spinner play a crucial role in enhancing localized suction force, significantly improving stone capture efficiency. As shown in **Figs. 2E and 2F**, the axial pressure distribution and the radial pressure distribution across the spinner cross-section at the location of maximum axial pressure drop reveal a pronounced and localized pressure drop toward the cavity in the spinner, reaching a peak negative pressure of –984 Pa. A comparison of the axial pressure drop between the slitted spinner and the non-slitted spinner demonstrates that the presence of slits induces a significantly greater pressure drop for much enhanced suction capability (**Fig. S1**). This pressure gradient is essential for effectively drawing stone fragments toward the spinner and ensuring their capture within the cavity. The design of the slits is carefully tailored to accommodate the size of kidney stone fragments. The slit width ($W_{slit}$) is smaller than the smallest targeted stone size whereas the inner diameter of the spinner ($ID$) needs to be larger than the maximum stone fragment size for stone entering. For example, when $W_{slit}$ = 0.4 mm and $ID$ = 3 mm, the spinner can effectively capture stone fragments ranging from 0.4 to 3 mm in size.

CFD simulations are performed to optimize the design of a spinner with an $ID$ of 3 mm and a $W_{slit}$ of 0.4 mm, aiming to maximize suction for optimal stone capture. The simulations specifically evaluate the effect of varying the number of slits ($n_s$) from 2 to 20 to determine the configuration that delivers the highest suction force and enhances stone capture efficiency (**Fig. 2G**). The results indicate that increasing in $n_s$ enhanced the maximum pressure drop along the spinner. Furthermore, the effect of total slit width (total $W_{slit}$) on pressure drop is investigated by maintaining a constant total $W_{slit}$ of 4 mm while varying the $n_s$ from 2 to 20, as shown in **Fig. 2G**. The maximum centerline pressure drop increases with rising $n_s$, peaking at 1 kPa when $n_s$ = 6 ($W_{slit}$ = 0.67 mm), and subsequently decreases gradually for $n_s$ >7. The simulated pressure drop comparison indicates that suction can be enhanced by maximizing the total $W_{slit}$ while keeping the $W_{slit}$ large enough for fluid exiting. In the study, considering fabricability and the minimum targeted fragment size of 0.4 mm, an optimal design with $n_s$ = 10 is adopted for the spinner to achieve effective stone capturing capability. Finally, the relationship between suction pressure and spinning frequency is also investigated using CFD simulation as shown in **Fig. 2H**. The maximum pressure drop is shown to increase monotonically with spinning frequency $f$. Specifically, the pressure drop rises from 64 Pa at 2.5k rpm to 1510 Pa at 12.5k rpm (**Fig. 2H**).

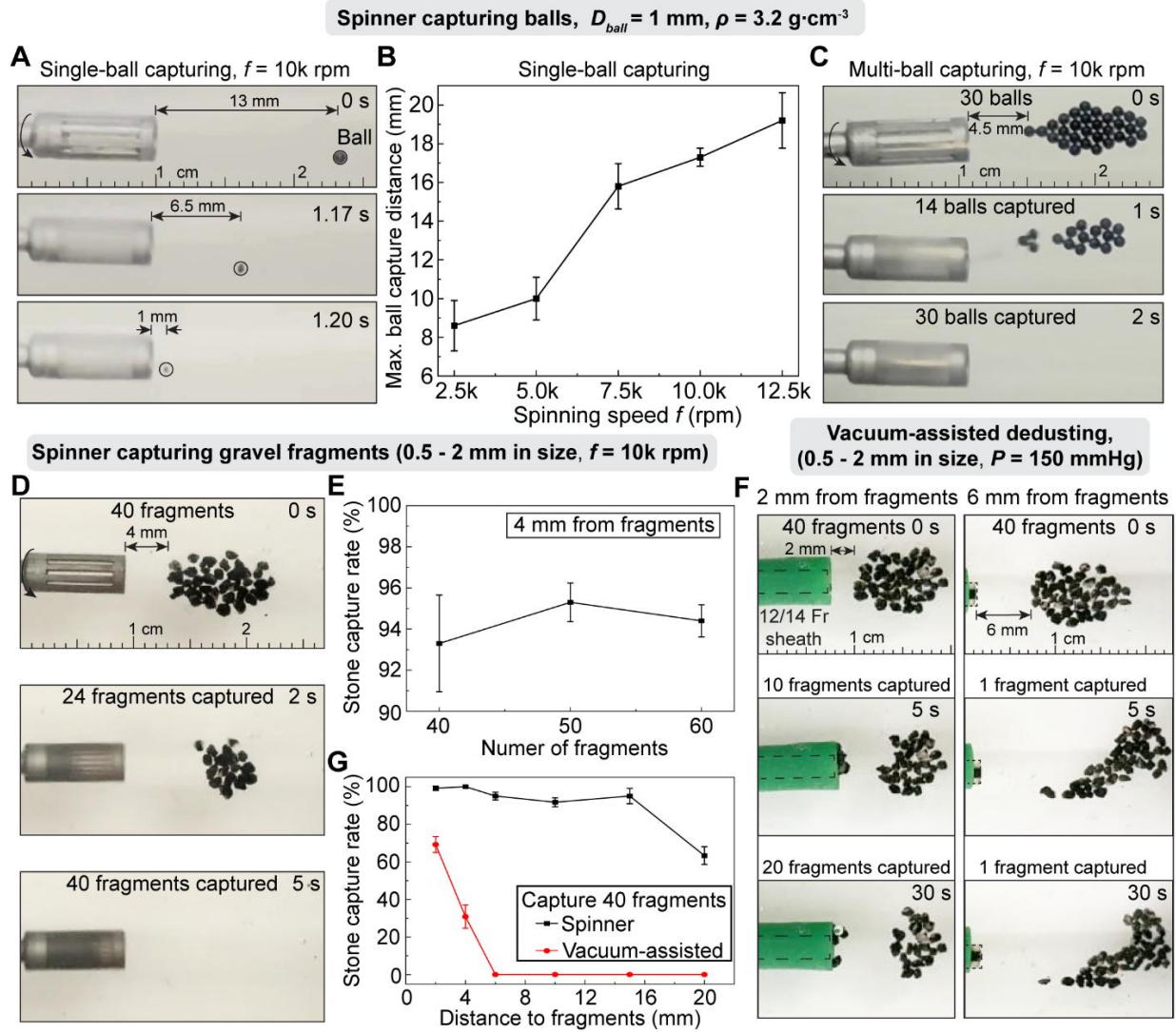

**Fig. 3. Evaluation of spinner performance in capturing balls and gravel fragments.** (**A**) Spinner at 10k rpm captures a single 1mm diameter ball (density = 3.2g/cm³) from a distance of 13 mm inside a 15mm diameter water-filled tube (approximating renal pelvis size) filled with water. All subsequent experiments are conducted in the same tube. (**B**) Maximum ball capture distance measured at spinner speeds of 2.5k, 5k, 7.5k, and 10k rpm. (**C**) Spinner at 10k rpm captures 30 balls placed 4.5 mm from its front surface (**D**) Spinner at 10k rpm captures 40 gravel fragments (0.5–2 mm) placed 4 mm from the front surface. (**E**) Capture performance at 10k rpm with 40, 50, and 60 gravel fragments (0.5–2 mm), each placed 4 mm away. (**F**) Vacuum-assisted dedusting test using a 12/14 Fr aspiration sheath with 150 mmHg suction and 60 ml/min irrigation through ureteroscope (dashed line) on 40 fragments placed 2 mm and 6 mm from the sheath opening. (**G**) Comparison of 40-fragment capture rates between the spinner and vacuum-assisted system at distances of 2, 4, 6, 10, 15, and 20 mm from the spinner or aspiration sheath.

**Characterization of spinner efficiency for capturing fragments**

The spinner kidney stone fragments capture ability is first experimentally demonstrated in **Fig. 3A** and characterized with spherical ceramic balls ($D_{ball}$ = 1 mm, $\rho$ = 3.2g/cm³, kidney stone has density around 1.7 to 3.2 g/cm³ (*23*)). The spinner rapidly draws and captures a ball from 13 mm distance away in 1.5s (**Movie S1**). **Fig. 3B** evaluates the maximum ball capturing distance with respect to spinning frequency, showing that increasing the spinning frequency from 2.5k to 12.5k rpm extends the maximum capture distance from 8.6 mm to 19.2 mm, a trend that correlates with the increased suction pressure observed in **Fig. 2H**. Additional experiments exploring capture distances for balls of different sizes and densities (encompassing the kidney stone density range) are presented in **Fig. S2**. Next, a multiple balls capture experiment further demonstrates the spinner's potential to collect a large number of stone fragments in a single ureteroscope pass, enhancing procedural efficiency and reducing retrieval time. In the test, 30 ceramic balls are placed 4.5 mm from the spinner (**Fig. 3C**) (**Movie S1**). At $f$ = 10k rpm, 14 out of 30 balls are captured within 1 s, and all 30 balls are successfully captured inside the spinner cavity in 2 s. (See Materials and Methods for experiment details).

In **Fig. 3D,** we demonstrate the capture of 40 irregular-shaped gravel fragments (density ~ 2 g/cm³, within kidney stone density range) ranging in size from 0.5 mm to 2 mm in single spin operation (See **Fig. S3** for fragment size distribution), showcasing its effectiveness in efficiently collecting multiple fragments in one run. 40 gravel fragments are placed 4 mm away from the spinner front. With 2 s of spinning, 24 out of 40 fragments are captured, and all 40 fragments are successfully collected in 5 s. The gravel fragment capture rate is examined against 40, 50, and 60 fragments placed 4 mm from the spinner surface in **Fig. 3E**. Overall, the fragment capture rate exceeds over 94% for all three cases. The spinner's fast collection of multiple fragments can drastically reduce the enormous ureteroscope passes required in traditional basketing or vacuum-assisted techniques, significantly enhancing procedural efficiency and reducing operation time.

In comparison with the vacuum-assisted dedusting technique that employs negative pressure to aspirate the fragments for removal, the spinner offers a more efficient and effective stone fragment removal utilizing its far-distance fragments capture ability. As shown in **Fig. 3F** and **Movie S2**, when 40 gravel fragments are placed 2 mm from a 12/14 Fr aspiration sheath (see Materials and Methods for the operational procedure), the vacuum-assisted technique, operating at 150 mmHg pressure with irrigation through the ureteroscope at 60 ml/min(*13, 19*), removes fragments located 2–3 mm from the sheath within 5 s. However, due to its significantly limited effective working distance, 30 stones remain uncollected; after 30 s, 20 fragments still remain as the sheath is blocked by the fragments and require ureteroscope retrieval for further aspiration. In contrast, under the same conditions, the spinner achieves a 100% capture rate in 10 s in a single run (**Fig. 3G**). Moreover, as the distance between the fragments and the devices increases from 4 to 15 mm, the capture rate of the vacuum-assisted technique drops to 0% starting at 6 mm distance. It is also observed that irrigation from the ureteroscope pushes fragments away from the sheath, making them impossible to aspirate (**Fig. 3F**). In contrast, the spinner maintains an impressive capture rate above 90% at 15 mm. These findings demonstrate that the spinner is significantly more effective at capturing stone fragments than vacuum-assisted techniques for three key reasons: i) it eliminates the need to chase fragments, as it actively draws stones into the spinner through its generated rotational flow field; ii) it greatly reduces the number of ureteroscope passes by capturing over 40 fragments in a single run; iii) it can draw fragments from a much greater distance. Together, these advantages can result in a substantial enhancement in operational efficiency.

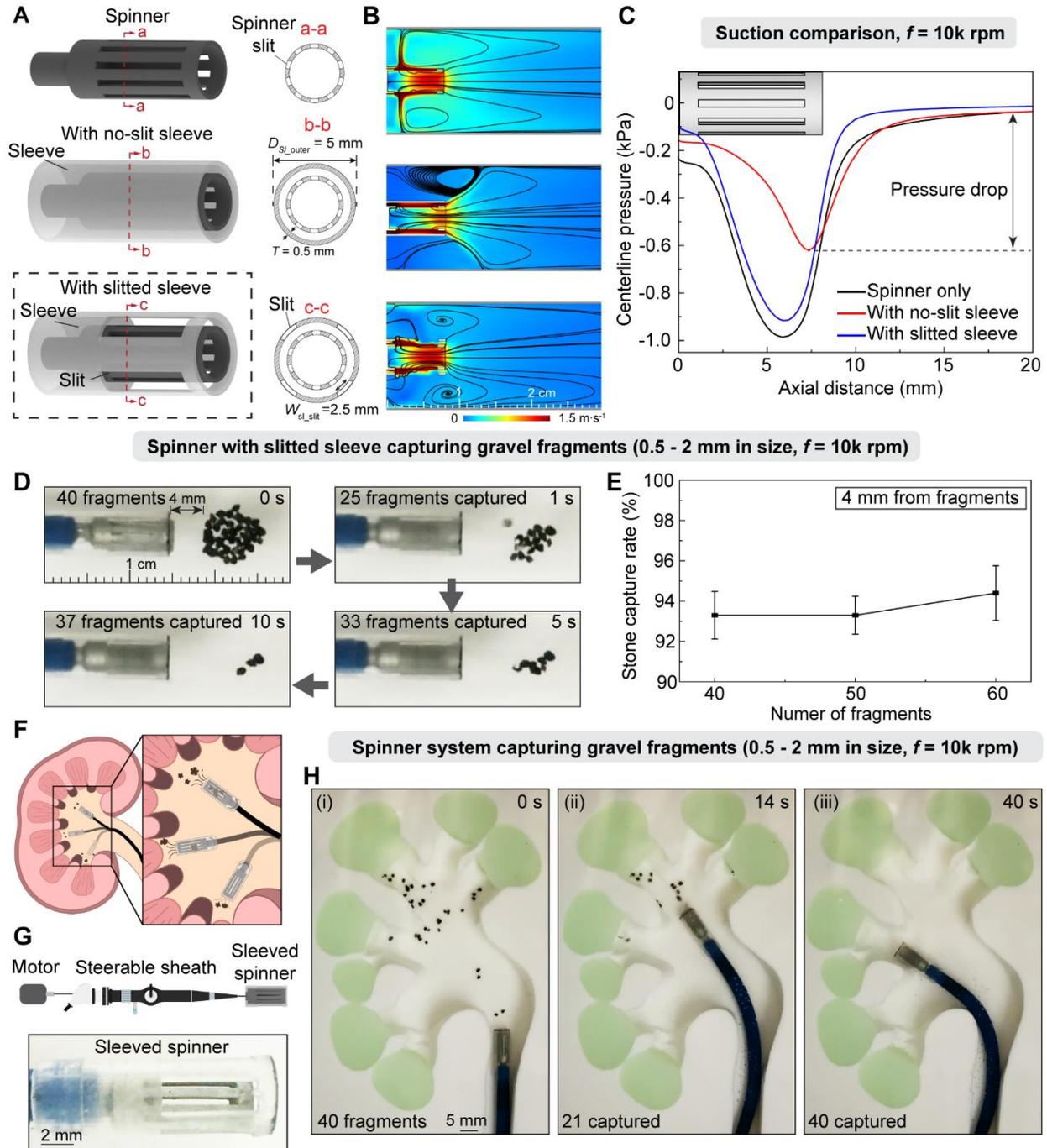

**Fig. 4. Sleeved spinner for enhanced safety and *in-vitro* efficiency test in kidney flow model.** (**A**) Schematics and cross-sectional views of three spinner designs: spinner alone, spinner with no-slit sleeve, and spinner with slitted sleeve. The sleeve has an outer diameter $OD_{sleeve}$= 5 mm and wall thickness $T$ = 0.5 mm, and slit width $W_{sl\_slit}$ = 2.5 mm. (**B**) Comparison of velocity contour with streamline produced on the longitudinal section of spinner designs inside 15 mm diameter tube. (**C**) Centerline pressure distribution along the spinner axis for the three designs. (**D**) Demonstration of slitted sleeve spinner 10k rpm capturing gravel fragments (0.5-2 mm in size) placed 4 mm away. (**E**) Capture rates when attempting to collect 40, 50, and 60 gravel fragments from 4 mm away. (**F**) Schematic of the spinner system integrated with a steerable sheath, operating

in a kidney flow model for fragment removal. (**G**) Schematic and photo of the distal end of the sleeved spinner integrated with a steerable sheath. (**H**) In vitro demonstration of fragment capture in the kidney flow model. (i) Initial placement of 40 gravel fragments in the upper pole. (ii) 21 fragments captured after 14 s of operation at 10k rpm. (iii) Complete capture of all 40 fragments within 40 s.

**Sleeved spinner for enhanced safety and *in vitro* efficiency in a kidney flow model**

To improve operational safety, a stationary slitted sleeve is integrated onto the spinner. This sleeve physically separates the spinning component from kidney tissue during operation. It is a thin, hollow, non-rotating cylinder with slits, designed to house the spinner (**Fig. 4A**, see **Fig. S4** for an image of the spinner system). The sleeve has an outer diameter ($D_{sl\_outer}$) of 5 mm, wall thickness ($T$) of 0.5 mm. Each slit on sleeve has slit width ($W_{sl\_slit}$) of 2.5 mm. The slits are crucial for maintaining circulation flow generated by the spinner. They act as outlets for fluid exiting through the spinner's slits, which facilitate strong localized suction by enabling a significant pressure drop inside the spinner cavity. To illustrate their importance, CFD analysis is conducted on three designs: spinner only, spinner with a no-slit sleeve, and spinner with a slitted sleeve. As shown in **Fig. 4B,** the no-slit sleeve obstructs fluid exit, disrupting circulation and reducing pressure drop to only 60% compared to the spinner only design (**Fig. 4C**). This results in weaker suction and shifts the pressure drop toward the front of the spinner, which reduces the likelihood of fragment capture. In contrast, the slitted sleeve allows fluid to exit through the sleeve slits, restoring circulation and bringing the pressure drop back to levels seen in the spinner-only design.

With the slitted sleeve, the fragments capturing performance of the spinner system is evaluated (**Fig. 4D** and **Movie S3**, see the Materials and Methods for sleeve manufacturing details). In this experiment, 40 gravel fragments (0.5-2 mm in size) are placed 4 mm in front of the spinner. The spinner and sleeve fronts are aligned to optimize capturing efficiency. After 1 s of spinning, 25 fragments are collected in the spinner cavity, increasing to 33 after 5 s, and 37 after 10 s. Experiments comparing the spinner's performance with and without sleeve slits demonstrate the critical role of the sleeve slits in maintaining effective suction (**Fig. S5**). The capture rate of the sleeved spinner is further examined by attempting to capture 40, 50, and 60 gravel fragments placed 4 mm away from the spinner. Additional tests with 40, 50, and 60 fragments showed consistent capture rates of 93.3%, 93.3%, and 94.4% within 20 seconds (**Fig. 4E**), demonstrating high efficiency across different loads.

An *in vitro* kidney flow model is used to assess the spinner's performance in a clinical-like setting (**Fig. 4F**) to simulate post-laser lithotripsy conditions. The spinner system is integrated with a steerable sheath with the spinner and slitted sleeve mounted at the distal end (**Fig. 4G**). A flexible shaft driven by a motor enables spinner rotation, and the steerable sheath navigates the system to target fragments placed in different kidney regions (See Material and Methods for setup details).

Forty gravel fragments (0.5-2 mm) are randomly distributed in the upper pole of the model (**Fig. 4H(i)**). The spinner at 10k rpm generates flow that rapidly dislodges and draws fragments into its cavity (**Movie S4**). After 13 s, 23 fragments are collected (**Fig. 4H(ii)**). After 40 s, all 40 fragments are removed, achieving a 100% capture rate. Unlike basketing, which requires direct targeting, or vacuum-assisted methods, which demand close proximity, the spinner can capture fragments across a wider area and from a distance. It also collects many fragments in a single pass, reducing the need for repeated ureteroscope insertions. This significantly improves procedural efficiency and minimizes trauma. The *in-vitro* kidney flow model demonstration highlights the potential of the spinner system to enhance post-lithotripsy stone clearance. The spinner's high fragment capture efficiency and rapid removal rate represent a significant advancement in RIRS, offering the potential to dramatically reduce procedure time and improving overall treatment outcomes compared to existing methods.

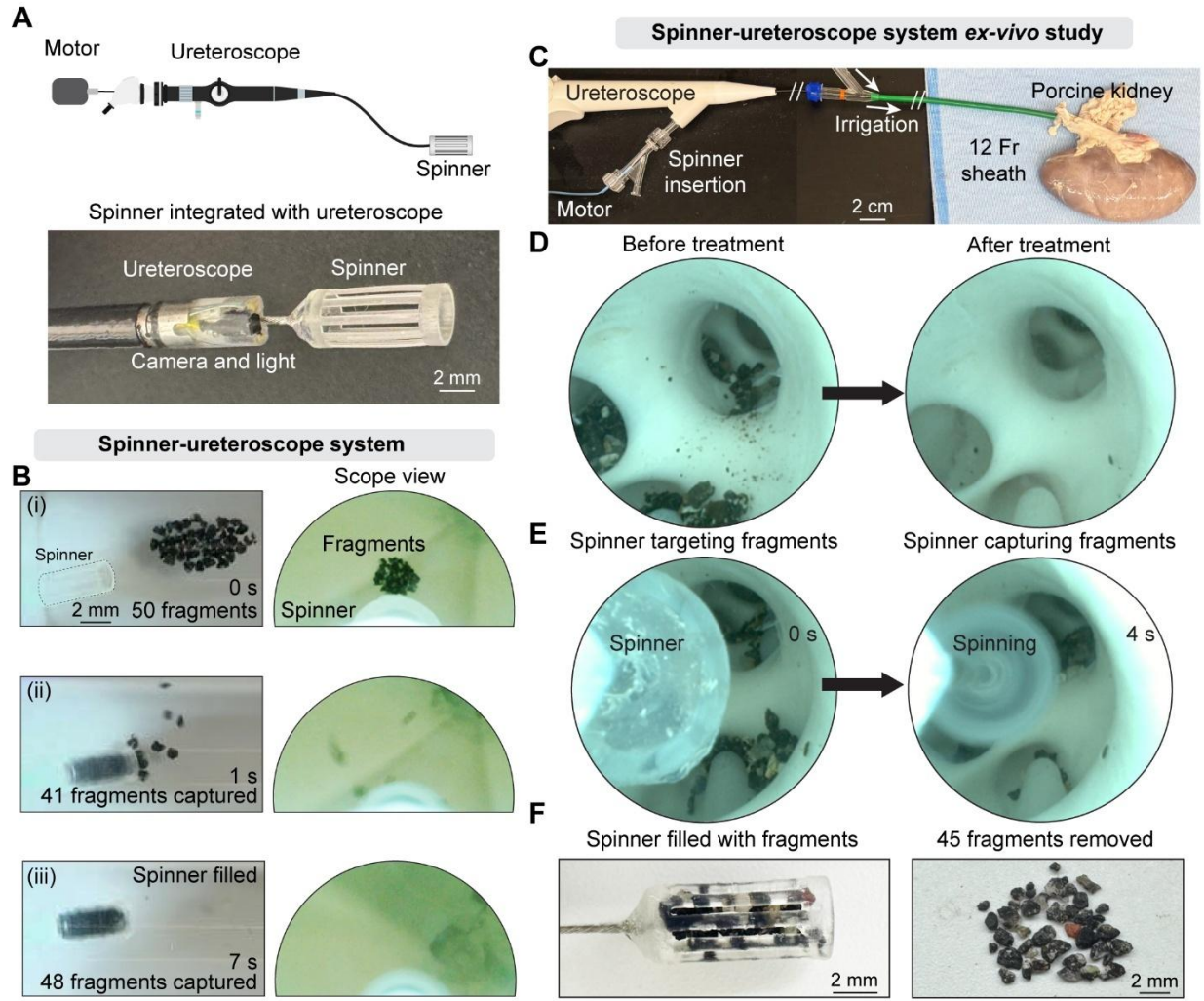

**Fig. 5.** *Ex-vivo* **fragment removal in porcine kidney using the spinner-ureteroscope system.** (**A**) Schematic and photo showing the integration of the spinner with a ureteroscope. (**B**) Spinner-ureteroscope system capturing 50 gravel fragments (0.5-2mm in size) at 10k rpm. (i) 50 fragments are placed distal to the spinner. capture is visualized via both top-down and ureteroscopic views. (ii) Within 1 s, 41 fragments are captured, demonstrating rapid suction. (iii) After 7 s, 48 fragments are captured, indicating near-complete removal. (**C**) Experimental setup of *ex-vivo* fragment removal in the porcine kidney model. (**D**) Scope views of before-and-after treatment comparison of the kidney showing near-complete fragment removal. (**E**) Scope views of a single operation showing large volume of fragments captured within 4 seconds of spinning at 10k rpm. (**F**) Post-operation image of the spinner and captured fragments from (E), filled with 45 captured fragments.

### *Ex-vivo* fragment removal in porcine kidney using the spinner-ureteroscope system

Fragment removal using the spinner device is performed in a porcine kidney to assess its performance within an intrarenal environment. For *ex-vivo* operation, the spinner is integrated with a ureteroscope, allowing real-time visualization of stone fragments inside the kidney. The spinner is positioned distal to the ureteroscope and driven by a motorized flexible shaft inserted through the ureteroscope working channel (**Fig. 5A**). Initial validation of the spinner–ureteroscope system is carried out in a tubular setup to demonstrate its fragment capture and visualization capabilities (**Fig. 5B** and **Movie S4**). Before operation, 50 fragments (0.5-2mm in size) are placed distal to the spinner and are clearly visualized via the ureteroscope (**Fig. 5B(i)**). within only 1 s of spinning at 10k rpm, the spinner rapidly captures 41 out of the 50 fragments, nearly filling its cavity (**Fig. 5B(ii)**). The ureteroscopic view shows that the fragments are dislodged and drawn towards the spinner. Ultimately, the spinner cavity is filled up by 48 out of 50 fragments in 7s (**Fig. 5B(iii)**), corresponding to a capture rate of 96%. The ureteroscopic view confirms the cleared field.

The spinner-ureteroscope system is then tested in an *ex-vivo* porcine kidney model (**Fig. 5C**). A 12/14 Fr flexible access sheath is introduced through the ureter to access the renal pelvis. To simulate post-laser lithotripsy conditions, gravel fragments (0.5-2 mm in size) are introduced through the sheath into the renal pelvis and kidney upper pole regions of the kidney (See Materials and Methods for experiment details), with the fragment distribution identified by ureteroscopic inspection (**Fig. 5Di**). The spinner system is then inserted through the access sheath and navigated to the renal pelvis for fragment removal. Continuous irrigation is applied through the sheath to maintain intrarenal pressure and ensure a clear visibility for operation. After a couple of passes, fragments in both the renal pelvis and upper pole are cleared (**Fig. 5Dii**). One representative operation is shown in **Fig. 5E** and **Movie S4**. At time 0 s, the spinner-ureteroscope system navigates to the targeted fragments located in the renal pelvis and upper pole. Within 4 s of spinning at 10k rpm, the spinner is retracted into the sheath and withdrawn from the kidney, completing one pass of fragment removal. In this single pass, 45 fragments fill up the spinner cavity (**Fig. 5F**). The *ex-vivo* study demonstrates the high efficiency of the spinner system for rapid stone fragment removal, which is expected to significantly enhance procedural efficiency by reducing operation time.

**Discussion**

In this work, we present a novel spinner device that enables ultra-efficient kidney stone fragment removal through spinning-induced localized suction. The spinner's uniquely engineered geometry generates a three-dimensional circulating and spiral flow field that effectively dislodges and draws fragments into its cavity. Its design is optimized via computational fluid dynamics to maximize suction performance and enhance fragment capture efficiency. The spinner's ability to capture a large volume of stone fragments across varied distances is systematically evaluated and compared with vacuum-assisted dedusting techniques, demonstrating superior performance in both capture rate and capture distance over 20mm. Its operational safety is further improved by incorporating a slitted sleeve, which ensures effective fragment collection while preventing tissue contact. The spinner's rapid and near-complete fragment removal capability is validated by in-vitro tests in a kidney flow model. Ex-vivo testing of the integrated spinner–ureteroscope system in a porcine kidney confirms its high efficiency by showing the capturing of 45 fragments in just 4 seconds during a single pass and achieving complete fragment clearance within a few passes. This high capture capacity is expected to dramatically enhance procedural efficiency by reducing operative time, minimizing ureteroscope passes, and improving stone-free rates.

    Encouraged by the promising results, future work is needed on optimizing the integration of the spinner with the ureteroscope to improve navigation, visualization, and steerability. This can be achieved by engineering a ureteroscope with an enlarged working channel capable of fully housing the spinner, thereby preventing direct contact between the spinning component and intrarenal tissue. These advancements are expected to further enhance procedural safety and efficiency, positioning the spinner as a next-generation technology for kidney stone fragment removal with the potential to significantly improve surgical outcomes.

**Materials and Methods**

*Spinner and sleeve manufacturing*

Spinner was printed using a commercialized stereolithography printer Form 3+ (Formlabs Inc., USA) with Formlabs Durable resin with a printing resolution of 50 μm.

*Maximum capturing distance testing*

The fragment capturing tests were conducted inside a 15 mm inner diameter, 10 cm long tube sealed at the distal end. Single balls made of different materials, zirconia ($ZrO_2$), silicon nitride ($Si_3N_4$), and polytetrafluoroethylene (PTFE), and in two sizes (1 mm and 2 mm) were tested. For each trial, a ball was placed inside the tube, and the spinner was inserted behind it. The initial distance between the spinner and the ball was set to 2 mm and gradually increased until the spinner could no longer capture the ball, therefore determining the maximum capture distance. Tests were repeated multiple times under varying spinning speed (2.5k, 7.5k, 10k, and 12.5k rpm) and across different ball densities and sizes.

*Multi-Ball capturing testing*

The multi-ball capturing tests were conducted using the same setup as the maximum capture distance tests, with the tube submerged in soapy water to reduce adhesion between the balls. Multiple balls were introduced into the tube using a pipette, forming a cluster. The spinner was then inserted and advanced to a preset distance from the first ball. Once in position, the spinner was activated at 10k rpm to capture the balls.

*Gravel fragments capturing testing*

Gravel fragments were first imaged to determine their size distribution using ImageJ software prior to testing. The fragments were then introduced into the same tube setup used in the multi-ball capturing tests, with a pipette used for positioning. Once inside the tube, the fragments were pushed together to form a cluster. The spinner was then activated at 10k rpm and operated for approximately 30 seconds. After the test, the spinner was removed, and the captured fragments were collected and counted.

*Vacuum-assisted gravel fragments removal*

Gravel fragments were placed inside a 15 mm inner diameter, 10 cm long tube capped at the distal end. Irrigation was delivered through a ureteroscope (2.8 mm in diameter) at a flow rate of 60 mL/min. The ureteroscope was inserted through a 12/14 Fr flexible access sheath, extending to the distal end of the sheath. The proximal end of the sheath was connected to a pump generating an aspiration pressure of 150 mmHg. The sheath–ureteroscope assembly was then placed into the 15 mm tube and advanced to a preset distance from the gravel fragments. Aspiration was activated first, followed by irrigation. During the procedure, the sheath and ureteroscope were held stationary. After the test, the number of captured fragments was counted, and the capture rate was calculated.

*In-vitro fragments removal in kidney flow model*

The kidney flow model was 3D-printed using polylactic acid (PLA). To replicate the anatomical structure, the renal medulla region—where stone fragments are present—was filled with dyed Dragon Skin 20 silicone. For testing, the model was submerged in water, and 40 gravel fragments were placed near the upper pole section. The spinner was integrated with an 8 Fr steerable sheath to allow maneuverability within the model. Once inserted into the flow model, the spinner was activated at 10k rpm and navigated to capture and remove the stone fragments.

*In-vitro fragments removal in porcine kidney*

Porcine kidneys were purchased from Animal Technologies. A 12/14 Fr flexible sheath was utilized to access the kidney through ureter. Gravel fragments were loaded into the kidney through the sheath. Irrigation was set to be running through the sheath to keep the kidney pressurized to prevent kidney collapsing. Before spinner-ureteroscope testing, the kidney was examined by an ureteroscope to check gravel fragments' location inside the kidney. Spinner was integrated with a standard ureteroscope (2.8 mm in diameter) to assemble the spinner-ureteroscope system, which was inserted through the sheath and into the kidney. The spinner was set to spin at 10k rpm to capture gravel fragments and pulled back while spinning to prevent fragments from falling off during the system retrieval. The captured fragments were counted.

*Computational fluid dynamics (CFD) simulations*

In this study, CFD simulations were performed using COMSOL Multiphysics 6.1 (COMSOL Inc., USA) to qualitatively evaluate spinner designs' effectiveness in generating suction. The simulation utilized the turbulent k-epsilon model and modeled the spinning motion through the frozen rotor method. The modeled environment mimicked the experiment condition where a 10 cm tube in length was caped at one end with the other end being open boundary condition.

**References**

1. T. Alelign, B. Petros, Kidney stone disease: an update on current concepts. *Advances in urology* **2018**, 3068365 (2018).
2. L. Frassetto, I. Kohlstadt, Treatment and prevention of kidney stones: an update. *American family physician* **84**, 1234-1242 (2011).
3. S. R. Khan *et al.*, Kidney stones. *Nature reviews Disease primers* **2**, 1-23 (2016).
4. D. Assimos *et al.*, Surgical management of stones: American urological association/endourological society guideline, PART I. *The Journal of urology* **196**, 1153-1160 (2016).
5. V. Gauhar *et al.*, Comparison and outcomes of dusting versus stone fragmentation and extraction in retrograde intrarenal surgery: results of a systematic review and meta-analysis. *Central European Journal of Urology* **75**, 317 (2022).
6. B. Van Cleynenbreugel, Ö. Kılıç, M. Akand, Retrograde intrarenal surgery for renal stones-Part 1. *Turkish journal of urology* **43**, 112 (2017).
7. G. Anan *et al.*, One-surgeon basketing technique for stone extraction during flexible ureteroscopy for urolithiasis: a comparison between novice and expert surgeons. *International Journal of Urology* **27**, 1072-1077 (2020).
8. S. L. Hecht, J. S. Wolf Jr, Techniques for holmium laser lithotripsy of intrarenal calculi. *Urology* **81**, 442-445 (2013).
9. E. Keller, V. De Coninck, S. Doizi, M. Daudon, O. Traxer, What is the exact definition of stone dust? An in vitro evaluation. *European Urology Supplements* **18**, e484 (2019).
10. D. H. Han, S. H. Jeon, Stone-breaking and retrieval strategy during retrograde intrarenal surgery. *Investigative and Clinical Urology* **57**, 229-230 (2016).
11. A. R. El-Nahas *et al.*, Dusting versus fragmentation for renal stones during flexible ureteroscopy. *Arab Journal of Urology* **17**, 138-142 (2019).
12. M. R. Humphreys *et al.*, Dusting versus basketing during ureteroscopy–which technique is more efficacious? A prospective multicenter trial from the EDGE research consortium. *The Journal of urology* **199**, 1272-1276 (2018).
13. C. Giulioni *et al.*, Experimental and clinical applications and outcomes of using different forms of suction in retrograde intrarenal surgery. Results from a systematic review. *Actas Urológicas Españolas (English Edition)* **48**, 57-70 (2024).
14. N. Liao *et al.*, A study comparing dusting to basketing for renal stones≤ 2 cm during flexible ureteroscopy. *International braz j urol* **49**, 194-201 (2023).
15. J. Liang *et al.*, Vacuum-assisted dedusting lithotripsy: a retrospective comparative study in high-risk patients with positive preoperative urine cultures. *World Journal of Urology* **43**, 128 (2025).
16. Z.-H. Wu *et al.*, Comparison of vacuum suction ureteroscopic laser lithotripsy and traditional ureteroscopic laser lithotripsy for impacted upper ureteral stones. *World journal of urology* **40**, 2347-2352 (2022).

17. V. Gauhar *et al.*, Technique, feasibility, utility, limitations, and future perspectives of a new technique of applying direct in-scope suction to improve outcomes of retrograde intrarenal surgery for stones. *Journal of Clinical Medicine* **11**, 5710 (2022).
18. Y. Chen *et al.*, A novel flexible vacuum-assisted ureteric access sheath in retrograde intrarenal surgery. *Bju International* **130**, 586 (2022).
19. J. Huang *et al.*, Vacuum-assisted dedusting lithotripsy in the treatment of kidney and proximal ureteral stones less than 3 cm in size. *World Journal of Urology* **41**, 3097-3103 (2023).
20. Z. Wen *et al.*, A systematic review and meta-analysis of outcomes between dusting and fragmentation in retrograde intrarenal surgery. *BMC urology* **23**, 113 (2023).
21. J. Uribarri, M. S. Oh, H. J. Carroll, The first kidney stone. *Annals of internal medicine* **111**, 1006-1009 (1989).
22. B. R. Matlaga *et al.*, Ureteroscopic laser lithotripsy: a review of dusting vs fragmentation with extraction. *Journal of endourology* **32**, 1-6 (2018).
23. J. W. Robinson, W. W. Roberts, A. J. Matzger, Kidney stone growth through the lens of Raman mapping. *Scientific Reports* **14**, 10834 (2024).

**Acknowledgement**

R. R. Zhao acknowledges the Stanford Terman Fellowship and Stanford Gabilan Fellowship.

**Author contributions**

Conceptualization: RRZ

Methodology: RRZ

Investigation: YC, JV, YS, RRZ

Visualization: YC, JV

Supervision: RRZ

Writing-original draft: YC, RRZ

Writing-review & editing: YC, RRZ

**Data and materials availability**: All data needed to evaluate the conclusion in the paper are present in the paper or in the Supplementary Material